\documentclass[prl,twocolumn,superscriptaddress,floatfix]{revtex4-1}
\usepackage{amsmath,amsfonts,amssymb,amsthm,graphics,graphicx,epsfig,bbm}
\usepackage[colorlinks=true,citecolor=blue,linkcolor=blue,urlcolor=blue]{hyperref}
\usepackage[usenames]{color}
\usepackage{graphicx}
\usepackage{subfigure}
\usepackage{epsfig}
\usepackage{bbold}
\usepackage{dcolumn}
\usepackage{bbm}
\usepackage{color}
\usepackage{verbatim}
\usepackage{epstopdf}
\usepackage{amssymb} 
\usepackage{amstext}
\usepackage{latexsym}
\usepackage{hyperref}
\usepackage{amsfonts}
\usepackage{psfrag}
\usepackage{xcolor}
\usepackage{times}
\usepackage{mathtools}
\usepackage{braket}
\usepackage{soul}

\begin{document}
\title{Protected State Transfer via an Approximate Quantum Adder}
\author{G. Gatti}
\affiliation{Departamento de Ciencias, Secci\'on F\'isica, Pontificia Universidad Cat\'olica del Per\'u, Apartado 1761, Lima, Peru}
\author{D. Barberena}
\affiliation{Departamento de Ciencias, Secci\'on F\'isica, Pontificia Universidad Cat\'olica del Per\'u, Apartado 1761, Lima, Peru}
\author{M. Sanz}
\email{mikel.sanz@ehu.eus}
\affiliation{Department of Physical Chemistry, University of the Basque Country UPV/EHU, Apartado 644, E-48080 Bilbao, Spain}
\author{E. Solano}
\affiliation{Department of Physical Chemistry, University of the Basque Country UPV/EHU, Apartado 644, E-48080 Bilbao, Spain}
\affiliation{IKERBASQUE, Basque Foundation for Science, Maria Diaz de Haro 3, 48011 Bilbao, Spain}

\begin{abstract} 

We propose a decoherence protected protocol for sending single photon quantum states through depolarizing channels. This protocol is implemented via an approximate quantum adder engineered through spontaneous parametric down converters, and shows higher success probability than distilled quantum teleportation protocols for distances below a threshold depending on the properties of the channel.

\end{abstract}

\date{\today}

\maketitle

\maketitle
Although decoherence has been described \cite{DecoDescribe1,DecoDescribe2}, measured \cite{DecoMeasure1,DecoMeasure2}, simulated \cite{DecoSimulate1,DecoSimulate2} and used for different quantum tasks \cite{DecoTask1,DecoTask2,DecoTask3}, it is often considered as one of the main setbacks of experimental setups based on quantum mechanics. This is evident in quantum computing, quantum simulations and quantum communication, which has triggered a race to find decoherence-resistant states and protocols, starting in the mid 1990s. Shor proposed one of the first quantum error correction codes for quantum computing \cite{Shor}, closely followed by the quantum distillation protocol for teleportation in noisy channels, proposed by Bennet \textit{et al.} in the framework of quantum communication \cite{Bennet}. The latter method is based on \textit{distilling} a few high fidelity Bell pairs from several low fidelity copies. Later, decoherence slowdown through quantum feedback was proposed by Vitali \textit{et al.} \cite{Vitali}, and the usage of noiseless (zero decoherence) subspaces was proposed by Zanardi \textit{et al.} \cite{Zanardi}. Quantum error correction codes based on topologically ordered states, such as a spin-1/2 honeycomb lattice, have been proposed \cite{Honeycomb1,Honeycomb2} and experimentally demonstrated \cite{ExpHoneycomb}. Additionally, methods making use of weak measurements have emerged to protect quantum states from decoherence \cite{Weak,Weak2}. Recently, a cavity state transfer protocol assisted by temporal modes in a tailored waveguide was proposed in Ref. \cite{ZollerPST}. In all of these examples, we can distinguish error correction from decoherence protection methods. Within the latter group, quantum optics is a natural choice for researching protected state transmission against decoherence.

In the early 1990s, interference between coherently pumped down-converters emerged as a powerful tool in quantum optics. Particularly, the second-order spontaneous parametric down-conversion interference device (SPDC interference) \cite{Mandel,SPDCHeuer} was proven to be key in fundamentals of quantum mechanics due to its nonlocality \cite{ZeilingerOld,ZeilingerOld2,ZeilingerOld3}. A variant of this array was exploited for ghost imaging without requiring coincidences \cite{Zeilinger}.

In the search for decoherence-protected systems, we will link SPDC interferometers with the theoretical construction of quantum adders \cite{Qadder}. A quantum adder is defined as a transformation in which the output is a superposition of two arbitrary unknown input states, previously codified in two different Hilbert spaces \cite{Qadder,Qadder2}. This transformation is forbidden by quantum mechanics, but it can still be achieved by postselection for states which are nonorthogonal to a reference state \cite{ExpQadder}. Despite the experimental realization, no practical application has been found for probabilistic quantum adders yet.

In this Letter, we propose a protocol for sending single photon quantum states protected against depolarizing channels. The setup is a direct application of a probabilistic quantum adder based on a second order SPDC interferometer. Our protocol consists in encoding a qubit, initially written in the polarization of a single photon, into the Fock space of two different paths. This encoded qubit is sent through decoherence channels and its original information is afterwards reconstructed by means of another nonlinear crystal. Our success probability for sending a qubit is not only higher than directly sending the state, but also than that obtained with distilled quantum teleportation for distances below a threshold depending on the properties of the channel.

\paragraph{State preparation via quantum adder. --}

A quantum adder \cite{Qadder} consists in the transformation $\ket{\alpha}\ket{\psi_B}\ket{\psi_A}\rightarrow \ket{\alpha'}\ket{\chi} R\big(\ket{\psi_A}+\ket{\psi_B}\big)$, where $\ket{\psi_A}$ and $\ket{\psi_B}$ are arbitrary states initially in two different Hilbert spaces (for simplicity, we choose qubits), $R$ is a normalization constant, $\ket{\chi}$ is a state that may also depend on the input states, and $\ket{\alpha}$ and $\ket{\alpha'}$ are ancillas. This quantum adder is forbidden by quantum mechanics \cite{Qadder,Qadder2}.

In our setup, we will aim at a more general kind of quantum adder in which not only a sum is considered, but also different linear combinations of input polarization states. For the sake of simplicity, we fix one of these input states. The type of linear combination is controlled by a parameter $f$, which may depend on any of the initial states. This transformation is

\vspace{-0.5cm}
\begin{equation}
\ket{\alpha}\ket{\psi_B}\ket{\psi_A}\rightarrow \ket{\alpha'}\ket{\chi} R\big(\ket{\psi_A}+f \ket{C}\big)\mbox{,}
\label{QadderTran}
 \end{equation}

\noindent where $\ket{C}$ is our fixed input state.

\begin{figure}[h]
\centering
\includegraphics[scale=0.6]{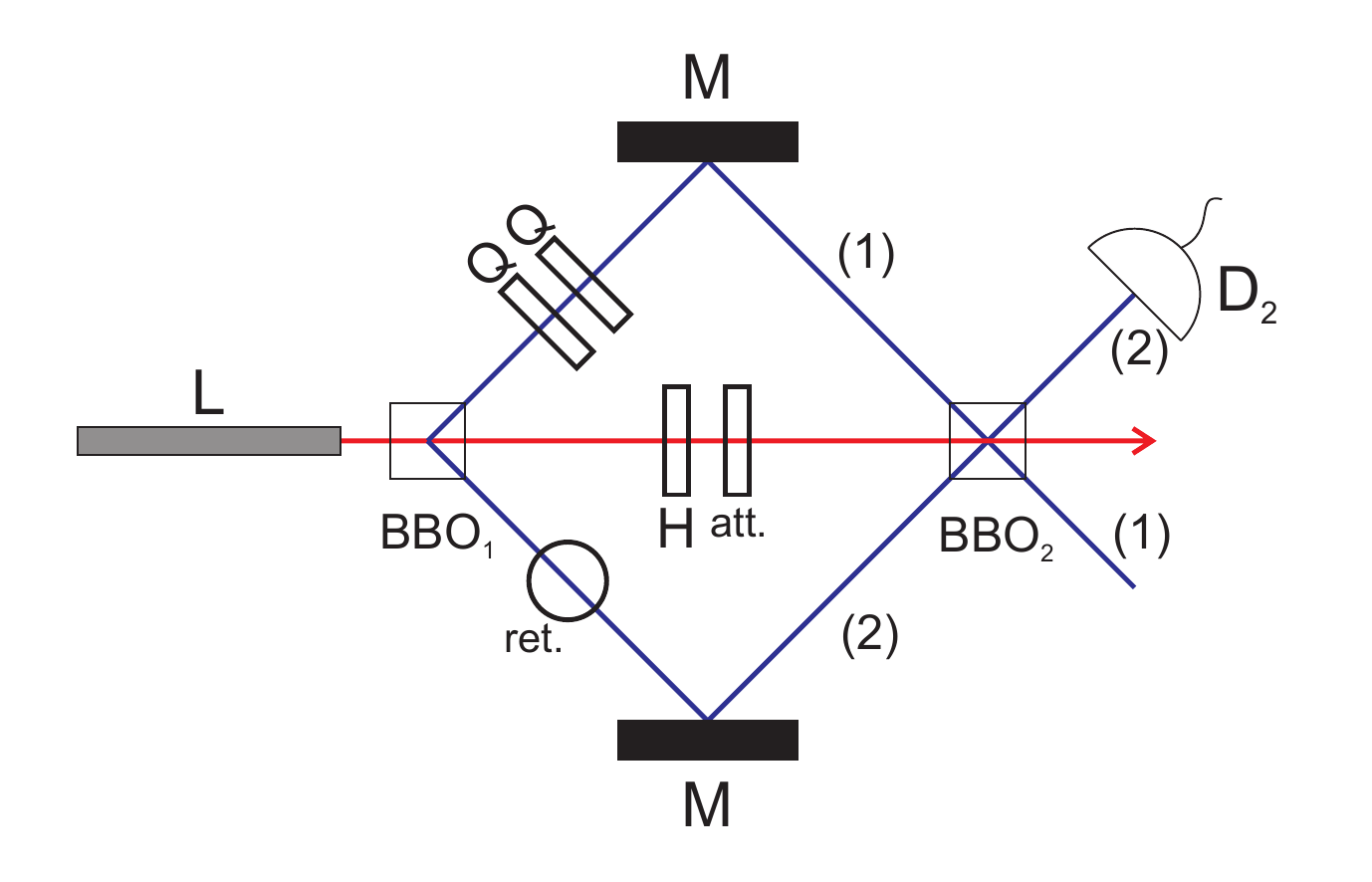}
\caption{A coherent laser (red line) pumps two type-1 beta-barium-borate crystals (\textit{BBO}), which, with probability amplitude $g$, emit pairs of single photons with polarization $\ket{V}$ in the specified paths (blue lines), via spontaneous parametric down-conversion (SPDC). The paths of the photons emitted by $\text{BBO}_1$ are aligned with mirrors so that they match with the paths that $\text{BBO}_2$ emissions (with the same frequency) would follow. \textit{Q} and \textit{H} are quarter and half waveplates, respectively. \textit{ret.} is a retarder, \textit{att.} is an intensity attenuator, \textit{M} are mirrors, and $D_2$ is a detector in path (2). Detections after $\text{BBO}_2$ show interference between probability amplitudes of pairs of photons emitted in $\text{BBO}_1$ and $\text{BBO}_2$. Consequently, this is not interference between paths (1) and (2).}
\label{doubleBBO1}
\end{figure}

To achieve this, let us consider a pumped non-linear crystal ($\text{BBO}_1$) where the paths of twin photons generated by spontaneous parametric down-conversion are aligned into a second crystal ($\text{BBO}_2$) (see Fig.~\ref{doubleBBO1}). As aforementioned, our setup is inspired in previous SPDC interferometers \cite{Mandel,ZeilingerOld}.

When twin photons are detected after $\text{BBO}_2$, it is indistinguishable which $\text{BBO}$ made the emission, and the uncertainty of the time of emission is sufficiently large to allow interference between both possibilities.

The two quarter-wave plates (\textit{Q}) are sufficient to rotate the polarization of path (1) photons in one of our emission possibilities, $\text{BBO}_1$, into an arbitrary chosen polarization $\ket{\psi_A}=a_1 \ket{H}+a_2\ket{V}$ \cite{DeZela}. Without loss of generality, we choose $a_2$ to be real. The retarder (\textit{ret.}) compensates for any extra phase introduced by this rotation and by any path difference in the experiment, and is set to add a relative phase $\phi$ between the coherent laser and the photons. The half-wave plate (\textit{H}) is set to rotate the pump beam, initially in horizontal polarization, into an arbitrary chosen polarization $\ket{\psi_B} = b_1 \ket{H} + b_2\ket{V}$, with $b_1$ and $b_2$ real, while the attenuator (\textit{att.}) is set to reduce the intensity of the coherent laser by a factor of $A$, with $0\leq A\leq 1$. Note that our protocol, after $\text{BBO}_2$, involves interference between the probability amplitudes of emitted photons by $\text{BBO}_1$ and $\text{BBO}_2$. Therefore, we are not dealing with interference between paths (1) and (2).

Let us address this in a more formal manner. The initial state of the system, before $\text{BBO}_1$, is $\ket{\alpha}_{0H} \ket{0}_{0V} (\ket{0,0}_{1}\ket{0,0}_{2})$, where $\ket{\alpha}_{0H} \ket{0}_{0V}$ indicates that the laser pump has a  horizontally polarized coherent state $\ket{\alpha}$ and a vertically polarized coherent state $\ket{0}$. Additionally, $\ket{n_H,n_V}_x$ indicates that there are $n_H$ ($n_V$) horizontally (vertically) polarized photons in path $x$, with $x=1,2$. Note that we are not using a two-level description for the single-photon polarization, since we will have zero-photon states in our description. We will, however, return to the two-level description of polarization later on.

For the interaction between our system and the BBOs, we use the effective Hamiltonian $H_{\text{BBO}}=g' (\hat{a}_{0H} \hat{a}_{1V}^\dagger \hat{a}_{2V}^\dagger+\hat{a}_{0H}^\dagger \hat{a}_{1V} \hat{a}_{2V})$ \cite{Walls}, where $\hat{a}_{x\,y}$ is the annihilation operator for path index $x$ (0, 1 or 2) and polarization $y$ (H or V), and $\hat{a}_{x\,y}^\dagger$ its respective creation operator. $g'$ is a crystal-dependent constant, which we assume to be real without loss of generality. We define $g=- \frac{g' t}{\hbar}$, where $t$ is the interaction time for the specific single-photon paths chosen. Note that $g$ is small because $t$ is of the order of the single-photon coherence length divided by the speed of light. We also consider $\alpha \ll g^{-1}$ for all coherent states $\ket{\alpha}$ considered in this setup, to allow us to keep only low orders.

Right before $\text{BBO}_2$, up to order $g^3$ in probability, we have the quantum state

\vspace{-0.5cm}
\begin{multline}
\ket{e^{i \phi}A \alpha b_1}_{0H} \ket{e^{i \phi} A \alpha b_2}_{0V}\\
\Big((1-\frac{g^2|\alpha|^2}{2})\ket{0,0}_{1}\ket{0,0}_{2}\\
+i g\alpha \big(a_1 \ket{1,0}_{1}+a_2 \ket{0,1}_{1}\big)\ket{0,1}_{2}\Big)\mbox{,}
\label{preBBO2}
\end{multline}

\noindent and after $\text{BBO}_2$, to same order, we have the state

\vspace{-0.5cm}
\begin{multline}
\ket{e^{i \phi} A \alpha b_1}_{0H} \ket{e^{i \phi} A \alpha b_2}_{0V}\\
\Big((1-\frac{g^2 |\alpha|^2}{2} (1+A^2 |b_1|^2+2 e^{-i \phi} a_2 {b_1} A)\ket{0,0}_{1}\ket{0,0}_{2}\\
+i g\alpha \big(a_1 \ket{1,0}_{1}+(a_2+e^{i \phi} A b_1) \ket{0,1}_{1}\big)\ket{0,1}_{2}\Big)\mbox{.}
\end{multline}

If we placed a detector in paths (1) or (2) just before $\text{BBO}_2$, the system state whenever a photon is received could be written using the $\{\ket{H},\ket{V}\}$  notation for polarization as

\vspace{-0.5cm}
\begin{multline}
\ket{e^{i \phi}A \alpha b_1}_{0H} \ket{e^{i \phi} A \alpha b_2}_{0V}\\
\big(a_1 \ket{H}_{1P}+a_2 \ket{V}_{1P}\big) \ket{V}_{2P}\mbox{,}
\label{AdderIn}
\end{multline}

\noindent since there would be no zero-photon states.

If we instead placed the detector after $\text{BBO}_2$, the state upon detection could be written as

\vspace{-0.5cm}
\begin{multline}
\ket{e^{i \phi} A \alpha b_1}_{0H} \ket{e^{i \phi} A \alpha b_2}_{0V}\\
R \big(a_1 \ket{H}_{1P}+(a_2+e^{i \phi} A b_1) \ket{V}_{1P}\big) \ket{V}_{2P}\mbox{,}
\label{AdderOut}
\end{multline}

\noindent where $R=\big(|a_1|^2+|a_2+e^{i \phi} A b_1|^2\big)^{-1}$.

Therefore, detected states before $\text{BBO}_2$ can be mapped via our probabilistic quantum adder into detected states after $\text{BBO}_2$. This effective transformation is given by Eq. \eqref{QadderTran}.

\paragraph{Protected state transfer. --}

In this section, we use the previous setup for a quantum state transfer protocol protected against decoherence. Alice wants to send an arbitrary qubit $\ket{\psi_A}=a_1 \ket{H}+a_2 \ket{V}$ to Bob and she creates this state in our previous setup by using the two \textit{Q} waveplates. Alice sets $b_1=1$, $A=a_2$ and $\phi=\pi$ with her \textit{H} waveplate, attenuator, and retarder. After $\text{BBO}_2$, the system state is

\vspace{-0.5cm}
\begin{multline}
\ket{S_1}=\ket{- a_2 \alpha}_{0H}\ket{0}_{0V}\\
\Big((1+\frac{g^2 {a_2}^2 |\alpha|^2}{2})(1-\frac{g^2|\alpha|^2}{2})\ket{0,0}_{1}\ket{0,0}_{2}\\
+i g\alpha a_1 \ket{1,0}_{1}\ket{0,1}_{2}\Big)\mbox{.}
\label{predeco}
\end{multline}

Note that a similar output would be obtained by setting $b_1=a_2$, $A=1$ and $\phi=\pi$.

Bob can not directly measure the state $\ket{\psi_A}$ from the current system state, but he can perfectly recover it by feeding the three signals into an additional nonlinear crystal, namely $\text{BBO}_3$ (compensating for path differences). Although it is technically easier to pump $\text{BBO}_3$ with the same coherent laser as the other crystals to avoid phase fluctuations between two different laser sources, for the purpose of minimizing transmitted resources, we will consider that it is pumped by an identical yet independent laser. This can be done because SPDC leaves coherent pump states invariant up to order $g^3$ in probability, and consequently BBO emissions are still indistinguishable when different pumps are used.

It could be argued that part of the information of the sent state $\ket{\psi_A}$ is in the coherent laser state that Bob needs to pump $\text{BBO}_3$. However, the coherent laser state right after $\text{BBO}_2$ can be easily fixed as $\ket{\alpha_2}$ for all possible states $\ket{\psi_A}$ (up to a zero-measure set) by regulating the initial intensity of the coherent laser accordingly. Taking the values of $b_1$ and $A$ into account, Alice would have to control $\ket{\alpha}$, so that $\ket{\alpha_2}$ would be constant. In this way, the coherent state needed to pump $\text{BBO}_3$ can be agreed beforehand, and only depends on the decoherence channel.

\begin{figure}[h]
\centering
\includegraphics[scale=0.48]{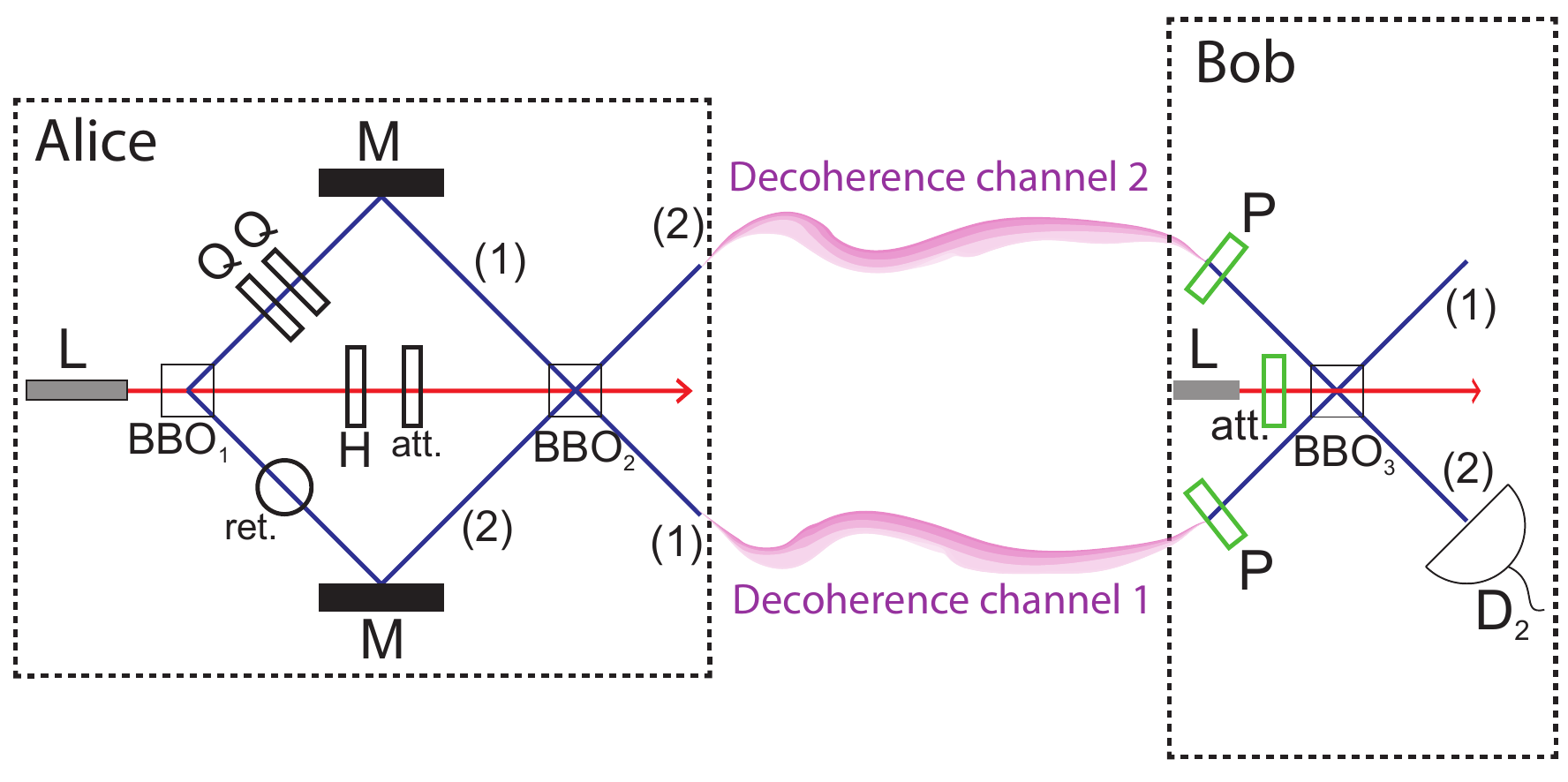}
\caption{Alice wants to send an arbitrary qubit $\ket{\psi_A}=a_1 \ket{H}+a_2 \ket{V}$ to Bob. She writes and encodes her qubit using our previous quantum adder setup. She then sets her H waveplate, attenuator and retarder so that $b_1=1$, $A=a_2$ and $\phi=\pi$, such that all single photons after $\text{BBO}_2$ have $\ket{\psi_A}$ independent polarizations. Afterwards, single photons are sent through decoherence channels and received by Bob. He projects them into their desired constant polarization, and attenuates his identical coherent laser to account for expected losses in the single photons. Finally, he feeds $\text{BBO}_3$ with those three signals, obtaining $\ket{\psi_A}$ in path (1) with a high success probability.}
\label{doubleBBO2}
\end{figure}

Before letting Bob reconstruct $\ket{\psi_A}$, we send our current state through decoherence channels affecting the single-photon paths. We show the behavior of our system under depolarizing and dephasing channels. The whole setup is depicted in Fig.~\ref{doubleBBO2}.

\paragraph{Depolarizing channel. --}

Let us compute the effect of the depolarizing channel on our setup and compare it with the success probability of both a direct transmission of the unencoded quantum state through the same noisy channel, and a similar setup using a distilled quantum teleportation protocol.

To calculate the decoherence caused by a depolarizing channel, we consider a three level system in path $x$ ($1$ or $2$), formed by any linear combination of the states $\{\ket{0,0}_{x},\ket{1,0}_{x},\ket{0,1}_{x}\}$. For simplicity, we will temporarily refer to these states as $\{\ket{0}_x,\ket{H}_x,\ket{V}_x\}$. The state $\ket{0}_x$ can not be dephased or flipped into $\ket{H}_x$ or $\ket{V}_x$ by depolarizing or dephasing decoherence channels (though the other states can dephase with respect to it).

Thus, for an initial density matrix

\vspace{-0.3cm}
\[ \rho_i = \left( \begin{array}{ccc}
a & x_1 & x_2 \\
{x}^*_1 & b & x_3 \\
{x}^*_2 & {x}^*_3 & c \end{array} \right)\]

\noindent written in our $\{\ket{0}_x,\ket{H}_x,\ket{V}_x\}$ basis, the density matrix after the depolarizing channel would be

\vspace{-0.3cm}
\[ \rho_f = \left( \begin{array}{ccc}
a & \sqrt{1-p}\, x_1 & \sqrt{1-p}\, x_2 \\
\sqrt{1-p}\, {x}^*_1 & \frac{(b+c) p}{2}+(1-p) b & (1-p)\, x_3 \\
\sqrt{1-p}\, {x}^*_2 & (1-p)\, {x}^*_3 & \frac{(b+c) p}{2}+(1-p) c \end{array} \right)\mbox{.}\]

Here, $p=1-e^{-\gamma L/c}$ is a parameter ranging between 0 and 1 that measures depolarization, with $\gamma$ is the depolarizing parameter, $L$ the channel distance and $c$ the speed of light. This transformation is applied on beams 1 and 2. Our initial state is given by Eq. \eqref{predeco}, where the single photons have $\ket{\psi_A}$ independent polarization. After the depolarizing channel, to correct the state as much as possible, we project path (1) and path (2) into horizontal and vertical polarization, respectively. Bob aligns the single photons into $\text{BBO}_3$ and pumps it with a (horizontally polarized) coherent laser $\ket{(1-\frac{p}{2})^2 \alpha_2}$, which is identical to the coherent state right after $\text{BBO}_2$, but attenuated by a factor of $(1-\frac{p}{2})^2$, the estimated intensity loss due to the decoherence channel and polarizers. After this, we calculate the success probability of our protocol, that is, the probability for path-(1) polarization states to be $\ket{\psi_A}$, and obtain

\vspace{-0.3cm}
\begin{equation}
P_{\text{PST}}=1-\frac{1}{4}\left(\frac{p/2}{1-p/2}\right)(1-\cos{4 \theta}) \mbox{,}
 \end{equation}

\noindent where we have parametrized $a_2=\sin{\theta}$.

\begin{figure}[h]
\centering
\includegraphics[scale=1.75]{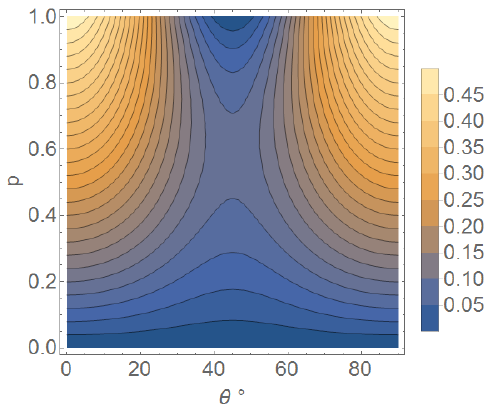}
\caption{Success probability enhancement (difference) of our protocol sending a qubit $\ket{\psi_A}=e^{i \phi} \cos{\theta} \ket{H}+\sin{\theta} \ket{V}$ through a depolarization channel, with respect to direct transmission. The degree of depolarization is parametrized by $p$, and $\theta$ is plotted between $0^{\circ}$ and $90^{\circ}$. The plot is symmetric respect to $\theta=0^{\circ}$.}
\label{versus1}
\end{figure}

If $\ket{\psi_A}$ is sent straightforwardly through the depolarizing channel, the success probability is $P_{\text{straight}}=1-\frac{p}{2}$. Our protocol has advantage in any point $\{p,\theta \}$ up to a zero-measure set ($p=0$, where both have perfect success, and $\{p=1,\theta=45^\circ \}$, where both have success probability 0.5). The enhancement of our protocol is shown in Fig.~\ref{versus1}.

Decoherence protected state transfer can also be realized with quantum distillation. Alice prepares $N$ Bell pairs and sends one party of each pair through the same noisy channel to Bob. This way, they share pairs of states $W_{F_0}$ that resemble Bell pairs, with probability $F_0=1-\frac{3p}{4}$. Quantum distillation consists in using LOCC between the two parties to reconstruct $m$ ($m<N$) pairs of states $W_{F}$, which resemble Bell pairs with probability $F$ ($F>F_0$). These states are then used in a teleportation protocol to transfer an arbitrary qubit between the two parties, so that $F$ is proportional to the success probability sending the qubit. We assume perfect classical communication.

Although there are several distillation protocols, we compare our proposal against the one with the best success probability without making use of additional pre-shared resources, within those proposed by Bennet \textit{et al.} \cite{Bennet}. In this protocol, every two states $W_{F_0}$ are distilled into one state $W_{F_1}$, where $F_1=1-\frac{2}{3}(1-F_0)$ to the lowest order in $(1-F_0)$, and there is a $\frac{2}{3}\sqrt{1-F_0}$ probability for this process to fail, when the pair of states $W_{F_0}$ are discarded. To same order, $k$ iterations of this procedure produce states $W_{F_k}$ with

\vspace{-0.4cm}
\begin{equation}
F_k=1-\left(\frac{2}{3}\right)^k (1-F_0)\mbox{.}
 \end{equation}

\begin{figure}[h]
\centering
\includegraphics[scale=1.75]{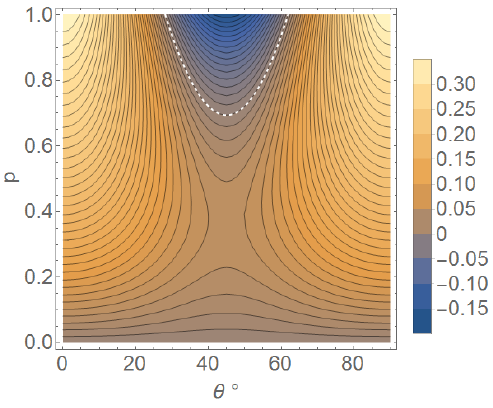}
\caption{Success probability enhancement (difference) of our protocol in sending a qubit $\ket{\psi_A}=e^{i \phi} \cos{\theta} \ket{H}+\sin{\theta} \ket{V}$ through a depolarization channel, with respect to distilled quantum teleportation protocol \cite{Bennet}. The region above the dashed line is the only region where the latter has higher success probability. The degree of depolarization is parametrized by $p$, and $\theta$ is plotted between $0^{\circ}$ and $90^{\circ}$. The plot is symmetric respect to $\theta=0^{\circ}$.}
\label{versus2}
\end{figure}

In both, our protected state transfer protocol and quantum teleportation with $k$ distillation iterations, the ratio between the expected number of states reconstructed by Bob and the number of states sent by Alice is a measure of the number of resources used. In distilled teleportation, it is $\frac{1}{2^k}\Big( 1-\frac{2}{3}\frac{1}{1-\sqrt{2/3}} \big(1-(\sqrt{2/3})^k \big) \sqrt{1-F_0}\Big)$, whereas in our protocol it is $(1-\frac{p}{2})^2$. We set these two quantities to be equal, in order to use the same amount of resources in both protocols, before comparing their respective success probabilities sending the qubit, as shown in Fig.~\ref{versus2}.

It is shown that, for $p\lesssim0.69$, our protocol has a better success probability that distilled quantum teleportation. Furthermore, when $p$ is greater than this threshold, distilled quantum teleportation only holds an advantage for $28^{\circ}\lesssim\theta\lesssim62^{\circ}$.

\paragraph{Dephasing channel. --}

Let us also compare the behavior of our protocol against a dephasing channel. Using a parameter $\beta_x$ between $0$ and $1$ to measure dephasing in single-photon path $x$, we note that the coherences between $\{\ket{H}_x,\ket{V}_x\}$ vanish quadratically faster than those between $\{\ket{0}_x,\ket{H}_x\}$ and between $\{\ket{0}_x,\ket{V}_x\}$. In our protocol, the coherences of the sent state are multiplied by a factor of $\sqrt{1-\beta_1}\sqrt{1-\beta_2}$, whereas in a straightforward transmission through path $x$ they are multiplied by a factor of $1-\beta_x$. The success probability of both protocols is thus the same when $\beta_1=\beta_2$.

Indeed, in our protocol, after reconstructing the state with $\text{BBO}_3$, the success probability is shown to be

\vspace{-0.5cm}
\begin{equation}
1-\frac{1}{2}\sin^2{(2 \theta)}(1-\sqrt{1-\beta_1}\sqrt{1-\beta_2})\mbox{,}
 \end{equation}

\noindent and the success probability for a polarization qubit straightforwardly sent through channel $x$ is
\begin{equation}
1-\frac{1}{2}\sin^2{(2 \theta)}\beta_x\mbox{.}
 \end{equation}

Summarizing, we propose a state transfer protocol with protection against depolarization. Our protocol shows success probability enhancement when compared against direct transmission and distilled quantum teleportation. Moreover, against symmetric dephasing channels, its success probability is not decreased with respect to direct transmission. Our setup is a practical application of a probabilistic quantum adder and is based on a spontaneous parametric down-conversion interferometer. This paves the way for novel long range quantum communication protocols through noisy channels and quantum information transfer.

We thank Ryan Sweke and Micha\l{} Oszmaniec for useful discussion and comments. We especially thank Francisco De Zela for helpful insights and suggestions. The authors acknowledge support from DGI-PUCP under Grant No. 2014-0064, CONCYTEC for fellowships, Spanish MINECO/FEDER FIS2015-69983-P and UPV/EHU UFI 11/55.

\end{document}